\documentclass[aps,prl,twocolumn,showpacs,superscriptaddress]{revtex4-1}

\usepackage{graphicx} 

\begin{document}

\title
{On the electronic structure of silicene on Ag(111): strong hybridization effects}

\author{S. Cahangirov}\email{seycah@gmail.com}
\affiliation{Nano-Bio Spectroscopy group, Dpto.~F\'isica de Materiales, Universidad del Pa\'is Vasco, Centro de F\'isica de Materiales CSIC-UPV/EHU-MPC and DIPC, Av.~Tolosa 72, E-20018 San Sebasti\'an, Spain}
\author{M. Audiffred}
\affiliation{Nano-Bio Spectroscopy group, Dpto.~F\'isica de Materiales, Universidad del Pa\'is Vasco, Centro de F\'isica de Materiales CSIC-UPV/EHU-MPC and DIPC, Av.~Tolosa 72, E-20018 San Sebasti\'an, Spain}
\affiliation{Departamento de Fisica Aplicada, Centro de Investigacion y de Estudios Avanzados, Unidad Merida, Km. 6 Antigua Carretera a Progreso, Apdo. Postal 73, Cordemex, 97310, Merida, Yuc., Mexico}
\author{P. Tang}
\affiliation{Nano-Bio Spectroscopy group, Dpto.~F\'isica de Materiales, Universidad del Pa\'is Vasco, Centro de F\'isica de Materiales CSIC-UPV/EHU-MPC and DIPC, Av.~Tolosa 72, E-20018 San Sebasti\'an, Spain}
\affiliation{Department of Physics and State Key Laboratory of Low-Dimensional Quantum Physics, Tsinghua University, Beijing 100084, People's Republic of China}
\affiliation{Institute for Advanced Study, Tsinghua University, Beijing 100084, People's Republic of China}
\author{A. Iacomino}
\affiliation{Nano-Bio Spectroscopy group, Dpto.~F\'isica de Materiales, Universidad del Pa\'is Vasco, Centro de F\'isica de Materiales CSIC-UPV/EHU-MPC and DIPC, Av.~Tolosa 72, E-20018 San Sebasti\'an, Spain}
\author{W. Duan}
\affiliation{Department of Physics and State Key Laboratory of Low-Dimensional Quantum Physics, Tsinghua University, Beijing 100084, People's Republic of China}
\affiliation{Institute for Advanced Study, Tsinghua University, Beijing 100084, People's Republic of China}
\author{G. Merino}
\affiliation{Departamento de Fisica Aplicada, Centro de Investigacion y de Estudios Avanzados, Unidad Merida, Km. 6 Antigua Carretera a Progreso, Apdo. Postal 73, Cordemex, 97310, Merida, Yuc., Mexico}
\author{A. Rubio}\email{angel.rubio@ehu.es}
\affiliation{Nano-Bio Spectroscopy group, Dpto.~F\'isica de Materiales, Universidad del Pa\'is Vasco, Centro de F\'isica de Materiales CSIC-UPV/EHU-MPC and DIPC, Av.~Tolosa 72, E-20018 San Sebasti\'an, Spain}

\date{\today}

\begin{abstract}
The electronic structure of the recently synthesised (3$\times$3) reconstructed silicene on (4$\times$4) Ag(111) is investigated by first-principles calculations. New states emerge due to the strong hybridization between silicene and Ag. Analyzing the nature and composition of these hybridized states, we show that i) it is possible to clearly distinguish them from states coming from the Dirac cone of free-standing silicene or from the sp-bands of bulk Ag and ii) assign their contribution to the description of the linearly dispersing band observed in photoemission. Furthermore, we show that silicene atoms contribute to the Fermi level, which leads to similar STM patterns as observed below or above the Fermi level. Our findings are crucial for the proper interpretation of experimental observations.
\end{abstract}

 \maketitle

Silicon chemistry has been compared with that of carbon, but both experimental and theoretical studies have illustrated great differences between structure and reactivity of compounds of the two group-IV elements and silicene, the silicon equivalent of graphene \cite{graphene, novoselov}, is not an exception. Unlike carbon atoms in graphene, Si tends to adopt $sp^3$ hybridization over $sp^2$ which results in slightly buckled structure of silicene.\cite{takeda, durgun, lok, seymur} Despite this buckling, the free standing silicene structure has enough symmetry to preserve the linearly crossing bands around the Fermi level.\cite{lok, seymur} This makes electrons of silicene behave as massless Dirac Fermions as in graphene.\cite{novoselov}

Free standing silicene has not been synthesized yet. However, experiments show growth of nanoribbons \cite{kara, lelay, aufray, padova, depadova, paola, paolade} and two-dimensional (2D) monolayers \cite{lalmi, lin, vogt, biberian, diboride, feng, chen, enriquez} of silicene on substrates, usually composed of silver. Interestingly, there is a debate on the structural configuration of silicene on Ag(111). Superstructures with ($2 \sqrt{3} \times 2 \sqrt{3}$) \cite{lalmi}, (3$\times$3) \cite{lin, vogt}, ($\sqrt{3} \times \sqrt{3}$) \cite{feng,chen} and other \cite{biberian,enriquez} silicene reconstuctions on Ag surfaces have been experimentally reported.

Photoemission measurements show linearly dispersing bands for silicene nanoribbons \cite{padova, paolade} and 2D silicene sheets \cite{vogt} supported on Ag substrates. Nonetheless, this intriguing aspect was not theoretically addressed in detail, but just a simple explanation was provided that linked the observed linear bands to the theoretically predicted Dirac cone of free standing silicene. More precisely, Vogt \textit{et al.} \cite{vogt}, used these linear bands together with STM images as compelling evidence for monolayer silicene. By combination of calculated and measured STM images, they have nicely shown that silicene sheets are arranged so that (3$\times$3) silicene supercells coincide with (4$\times$4) Ag(111) supercells, where a Si-Si bond length of 2.3~\AA~is in good agreement with the theoretical calculations.\cite{seymur} The ARPES measurements show a linear band at the K point of the (1$\times$1) silicene unitcell, which starts 0.3~eV below the Fermi level and has a Fermi velocity of $v_F=1.3 \times 10^6$~m/s.  Here the (3$\times$3) reconstructed silicene unit cell is composed of two atomic planes with twelve Si atoms close to the Ag substrate and six Si atoms further apart. This structure is further confirmed by full structural optimization calculations and by comparing the calculated STM images with the experiment. The similarity between STM images at applied bias of -1.4~eV, -0.5~eV, and 0.6~eV was used to confirm that the origin of the observed patterns was of structural nature. However, the STM data around the Fermi level or the ARPES data over the whole Brillouin zone was not discussed. We note that a linear dispersion around the Fermi level was also observed in the band structure of ($\sqrt{3} \times \sqrt{3}$) reconstructed silicene.\cite{chen} However, the exact positioning of silicene atoms on silver is not well characterized experimentally or theoretically in these structures. Also the linear dispersion is inferred indirectly from scattering patterns measured by STM.

In this letter, we provide a simple and coherent interpretation of the main experimental observations for single-layer silicene supported on Ag. We extend the analysis of first-principles calculations performed on the structures reported by Vogt \textit{et al.} \cite{vogt}. We first show that the calculated STM around the Fermi level (0.0~eV) has similar pattern to those for -1.4~eV, -0.5~eV, and 0.6~eV, which is contributed by silicene, thus implying that silicene on Ag is metallic. We also show that the linear band at the K point of the (1$\times$1) silicene reported by the ARPES measurement \cite{vogt} can be attributed to a new state that emerge due to the strong hybridization between silicene and Ag. The character of these new states can be clearly analysed in detail by projecting  those states on specific atomic orbitals. Our results will trigger further experimental and theoretical works to consolidate our conclusions.

The first step of our calculations based on density functional theory (DFT) has been to determine the lattice parameter of bulk Ag.\cite{calculation} Then we put one layer of (3$\times$3) silicene on top of 11 layers of (4$\times$4) Ag(111) and optimized the structure by keeping the bottom 8 layers fixed at the bulk Ag parameters. The resulting structure has the same configuration as reported in Ref.\cite{vogt}. The calculated constant current STM images are depicted in Fig.1 (a). These images are mainly contributed by Si atoms that are further apart from the Ag layer. They form a hexagonal superstructure with three Si atoms at the corners. The image at 0.0~eV is similar to those at -1.4~eV, -0.5~eV, and 0.6~eV meaning that Si atoms do indeed contribute to the states at the Fermi level.

\begin{figure}
\includegraphics[width=8.5cm]{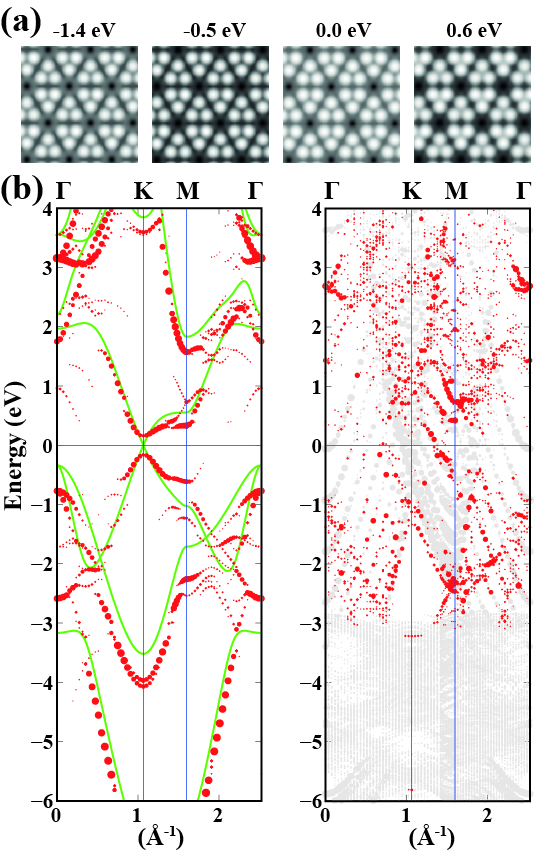}
\caption{(Color online) (a) The constant current STM images of reconstructed (3$\times$3) silicene on Ag substrate calculated at -1.4~eV, -0.5~eV, 0.0~eV, and 0.6~eV. (b) Left panel: bands of reconstructed (3$\times$3) silicene in the absence of Ag substrate (unsupported silicene) unfolded to BZ of (1$\times$1) silicene are shown by red dots. The radii of dots correspond to the weight of unfolding. The band structure of ideally buckled silicene is shown by green lines. Right panel: unfolded band structure of silicene on Ag. Red dots correspond to states with significant contribution from silicene.} \label{fig1}
\end{figure}

To understand the electronic structure of silicene on Ag and provide a microscopical description of the states contributing to the observed linear dispersion in the photoemission data of Ref.\cite{vogt}, we first look at the reconstructed silicene in the absence of the substrate which is referred as unsupported silicene throughout this letter. The corresponding band-structure unfolded to the (1$\times$1) unitcell of silicene is presented by red dots in the left panel of Fig.1 (b). Here the states in (3$\times$3) supercell are expanded in terms of the states in (1$\times$1) unitcell and the weight of contribution is represented by the radii of red dots.\cite{unfolding} The band structure of ideally buckled silicene in which one sublattice of Si atoms are at the bottom and other at the top is presented by green lines. For better comparison, the lattice constant and the buckling height of this structure is set to that of unsupported silicene. The symmetry of ideally buckled silicene preserves the linearly crossing bands at the K point. In the unsupported silicene, however, this symmetry is broken and the linear bands become parabolic with a 0.3~eV band-gap opening due to the structural modification induced by the Ag substrate (hybridization).

The unfolded band structure of our fully relaxed (3$\times$3) silicene on (4$\times$4) Ag substrate is shown in the right panel of Fig.1 (b). The $d$-bands of Ag are located in the energy window from 3~eV to 6~eV. Here we calculate the projection of each state on the atomic orbitals of silicene and mark with red dots the states in which the magnitude of projection is above some threshold. Comparing red dots in the left and right panel of Fig.1 (b), one can see significant change in the character of the silicene bands. In fact, it is hard to make a direct one to one correspondence between the states in these two panels. 

The $\sigma$-states around the $\Gamma$ point are shifted down by $\sim$0.5~eV while $\pi$-states around the K point are shifted down by $\sim$0.9~eV upon hybridization with Ag. This difference could be attributed to the larger overlap of the latter states with Ag substrate compared to the former ones. Note that, in the Brillouin zone (BZ) of (3$\times$3) silicene supercell, these two states fold to the same point. In the Fig.1 of the supplementary material, we analyse the charge densities projected on these states and show that the difference in the energy shift due to hybridization with Ag substrate results in the band inversion at the $\Gamma$ point of (3$\times$3) silicene. This also confirms that the shifts are mainly caused by the hybridization with Ag and not by the charge transfer from Ag to silicene, which is found to be 0.5 electron per (3$\times$3) supercell.\cite{oshiyama}

In Fig.1 (b) one can also see states at the Fermi level which has contribution from silicene. These are the states that give rise the STM patterns calculated at zero bias. In the supplementary material, we show that these states are folded to the K point of (3$\times$3) silicene and have significant contribution from $p_z$-states of six silicene atoms that are further apart from Ag substrate.

Recently (after submission of the present work), it was argued that, the slope and the energy range in which the experimentally  observed bands \cite{vogt} extend matches better to the sp-bands of Ag states than the Dirac cone of silicene.\cite{lin2} This argument, however, does not clarify why the linear bands dissapear when the measurement is performed in the absence of silicene. This fundamental piece of understanding  is provided below.

\begin{figure}
\includegraphics[width=8.5cm]{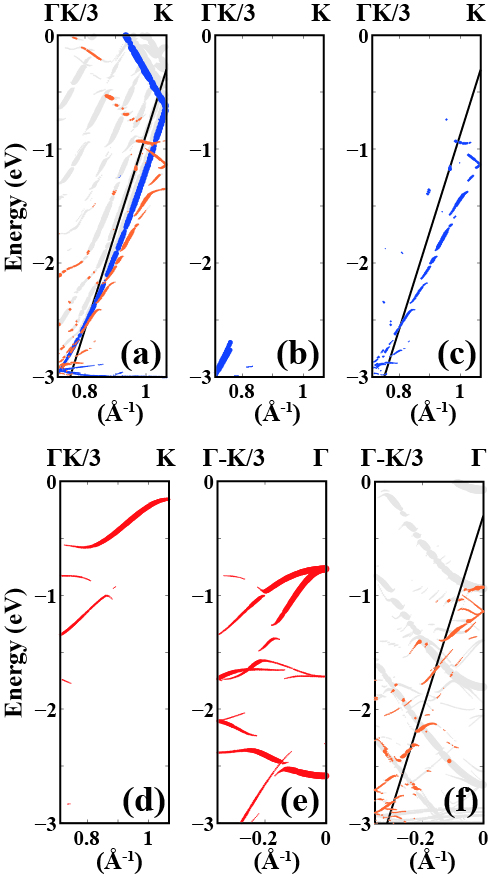}
\caption{(Color online) (a) Band structure of silicene with 11 Ag layers aroud K point. Blue and orange dots correspond to bulk Ag states and hybridized silicene/Ag states. Grey dots come from foldings of finite slab and can be ignored. Contribution of 3 Ag states at the surface in the (b) absence and (c) presence of silicene. (d) $\pi$-bands of unsupported silicene. (e) $\sigma$-bands of unsupported silicene around $\Gamma$ point. (f) Same as (a) around $\Gamma$ point. The solid black lines in (a), (c) and (f) are representational sketches of the experimental data.\cite{vogt} It is starting from -0.3~eV and has a slope corresponding to $v_F=1.3 \times 10^6$~m/s. It is recommended, however, to compare our data directly with the ARPES data.\cite{vogt}.} \label{fig2}
\end{figure}

The coherent interpretation of experiments requires analysing the band structure of (3$\times$3) silicene on 11 layers of (4$\times$4) Ag in a narrow window of the BZ and the energy range where the ARPES measurements with linear bands were reported.\cite{vogt}  In Fig.2 (a) we present the unfolded band structure of silicene on Ag system in this narrow window. Since the system includes 11 layers of Ag we assume that the 3 Ag layers in the middle can be considered as representative of the bulk Ag bands. We calculate the projection of each state on these 3 Ag layers in the middle and denote by blue dots the states that has magnitude of projection larger than some threshold. Then we calculate projections on 3 Ag layers underneath silicene and silicene itself and plot with orange dots the states having magnitude of projection above some threshold. These thresholds are defined such that the magnitude of contribution per atomic orbital is a constant. The same constant is also used in Fig.1 (b) and in the rest of Fig.2.

One can see in Fig.2 (a) that, the hybridization between silicene and Ag results in a state (shown in orange) which can be distinguished from the bulk Ag state (shown in blue). We confirmed that this hybridized state completely disappears when calculation is performed with 11 Ag layers without silicene. Note that, hybridization with Ag transforms the parabolic band structure with a gap into a gapless band structure with a quasi-linear dispersion that resembles a Dirac cone. Comparing with ARPES measurements\cite{vogt} one can see that the bulk Ag states indeed have an impressive overlap. However, the hybridized state has also considerable reminiscence.

To simulate the ARPES measurement which is sensitive to surface states\cite{vogt} we plot in Fig. 2 (b) and (c) the states contributed by 3 Ag layers underneath the silicene in the absence and presence of silicene, respectively. The threshold for projections is the same as in Fig. 2 (a). In this respect, it is very interesting that the bulk Ag states are absent in both plots. Moreover, the surface states of Ag have almost no contribution when the silicene is absent and the linear band appears when the silicene is present. This similarity with experiment and absence of Ag bulk contributions suggest that it is more reasonable to attribute the observed linear bands to the hybridized states of silicene and Ag rather than bulk Ag state, despite the likeness of the latter. Note that upon hybridization, the character of silicene states change significantly. In this respect, unsupported silicene states presented in Fig.2 (d) can be compared with hybridized states presented in Fig.2 (a). Despite this change, the charge densities of hybridized and unsupported silicene around the K point still show considerable resemblance (for more details see supplementary material).

We explore also the states near $\Gamma$ point because they can have contribution to experimental measurements due to the folding of (3$\times$3) silicene BZ.\cite{maria} Here we show $\sigma$-states of unsupported silicene and hybridized states around $\Gamma$ point in Fig.2 (e) and (f), respectively. The energy shift of states due to hybridization is less pronounced compared to $\pi$-states because $\sigma$-states have less overlap with Ag (see also supplementary material). Comparing Fig.2 (f) with Fig.2 (a) and (c) one can deduce that hybridized $\sigma$-states of silicene might also contribute to the experimental measurements.

Finally, we point out the improvements that our work bring and shortcomings that still remain on the interpretation of experimental results related with silicene on Ag. The hybridized state presented in Fig.2 (a) extends down to -3~eV where the Ag $d$-bands start, in accordance with experiments. The slope of the linear band presented in Fig.2 (c) corresponds to a velocity of $0.8 \times 10^6$~m/s that is considerably closer to the experimental value of $1.3 \times 10^6$~m/s than the Fermi velocity of free standing silicene ($v_F=0.53 \times 10^6$~m/s). We found that this Fermi velocity is increased to $v_F=0.71 \times 10^6$~m/s when $G_0W_0$ corrections are considered.\cite{gw,gwapl} Similar behaviour was found for graphene.\cite{gr1,gr2,gr3,gr4} The DFT slope of the hybridized state may be improved further by including those $G_0W_0$ many-body effects in calculations of silicene on Ag, as the states that we observe here comes mainly from the hybridization of the $sp$-bands of Si with the $sp$-bands Ag while $d$-bands of Ag contribute to the screening of electron-electron interactions at the Fermi level.

In summary, by means of first-principles calculations, we showed that the (3$\times$3) reconstructed silicene on Ag substrate has a metallic state at the Fermi level that shares a similar STM pattern as those for -1.4~eV, -0.5~eV, and 0.6~eV. This conclusion needs to be confirmed experimentally. Detailed band structure and charge density analysis reveal that the hybridization between silicene and Ag has substantial effect, leading to a change in the band order of some states. By calculating projections to specific atomic orbitals, we have clearly distinguished hybridized silicene/Ag states from bulk Ag states. We show that, hybridized states are localized at the surface and give rise to a linear band in the presence of silicene and disappear in the absence of silicene. This is in accordance with recent ARPES measurement which is sensitive to the surface states.\cite{vogt} Moreover, the hybridization between $\sigma$-states of silicene and Ag can also contribute to experimentally observed linear band due to the folding of (3$\times$3) silicene BZ.

In the light of our findings we conclude that it is not possible to attribute the experimentally observed linear bands solely to sp-bands of bulk Ag or to Dirac cones of silicene. We attribute the experimental bands to hybridized state localized at the surface emerging due to the interaction between silicene and Ag. This state has resemblance of experimental bands and appear only when silicene is present. The pronounced effects of hybridization with Ag substrate presented in this letter may play a crucial role in the interpretation of the linear bands observed also in single and multilayer silicene nanoribbons \cite{padova, paolade} and monolayer silicene structures with other type of reconstructions \cite{chen,chen2}. Their origin should be analysed in depth, instead of simply linking them to sp-bands of bulk Ag or to Dirac cones of silicene.

\textit{Acknowledgments} We acknowledge financial support from the European Research Council Advanced Grant DYNamo (ERC-2010-AdG -Proposal No. 267374) Spanish Grants (FIS2010-21282- C02-01 and PIB2010US-00652), Grupos Consolidados UPV/EHU del Gobierno Vasco (IT-578-13) and European Commission project CRONOS (280879-2 CRONOS CP-FP7). Computational time was granted by i2basque and BSC Red Espanola de Supercomputacion. MA thanks CONACYT for the Ph.D. fellowship and the financial support of project REA-FP7-IRSES TEMM1P (GA 295172). We thank Maria-Carmen Asensio, Jose Avila, Federico Iori and Pierluigi Cudazzo for fruitful discussions.


\begin{thebibliography}{15}
\bibitem{calculation} Calculations were carried out using the projector-augmented wave (PAW) pseudopotential method \cite{paw} as implemented in the VASP software.\cite{vasp} The generalized gradient approximation (GGA) in the Perdew-Burke-Ernzerhof form \cite{pbe} was used to include the exchange-correlation interactions. A plane wave energy cutoff of 300 eV was used. The vacuum spacing between image surfaces due to the periodic boundary condition is larger than 12.5~\AA. The forces on the relaxed atoms were converged to less than $10^{-3}$ eV/\AA. In the ionic relaxation calculations, the Brillouin zone was sampled by (3$\times$3$\times$1) k-points \cite{pack}, while the accurate charge density was calculated by (9$\times$9$\times$1) $\Gamma$ centered k-point grid. Bulk lattice constant of Ag was found to be 4.16~\AA. Buckling height of silicene on Ag was found to be 0.78~\AA~while the height difference between bottom Si and top Ag atoms is 2.30~\AA. The Tersoff-Hamann model was used for the simulation of STM images.\cite{stm} Corrections have been made to avoid fictitious dipole field due to the periodic slab images.\cite{dipole} A single-shot perturbative $G_0W_0$ calculation with default parameters provided by VASP \cite{gw} and (36$\times$36$\times$1) $\Gamma$ centered k-point grid was performed on top of DFT calculation of buckled (1$\times$1) silicene with a vacuum spacing of 15~\AA.
\bibitem{grey} States that do not belong to bulk Ag or hybridization with silicene (shown in grey) can be viewed as an artefact of using finite number of Ag slabs. In fact as the number of Ag slabs are increased these bands get more dense and converge to a grey background which can be ignored.
\end{thebibliography}
\end{document}


\title
{SUPPLEMENTAL-MATERIAL for "On the electronic structure of silicene on Ag substrate: strong hybridization effects"}

\author{S. Cahangirov} \email{seycah@gmail.com}
\affiliation{Nano-Bio Spectroscopy group, Dpto.~F\'isica de Materiales, Universidad del Pa\'is Vasco, Centro de F\'isica de Materiales CSIC-UPV/EHU-MPC and DIPC, Av.~Tolosa 72, E-20018 San Sebasti\'an, Spain}
\author{M. Audiffred}
\affiliation{Nano-Bio Spectroscopy group, Dpto.~F\'isica de Materiales, Universidad del Pa\'is Vasco, Centro de F\'isica de Materiales CSIC-UPV/EHU-MPC and DIPC, Av.~Tolosa 72, E-20018 San Sebasti\'an, Spain}
\affiliation{Departamento de Fisica Aplicada, Centro de Investigacion y de Estudios Avanzados, Unidad Merida, Km. 6 Antigua Carretera a Progreso, Apdo. Postal 73, Cordemex, 97310, Merida, Yuc., Mexico}
\author{P. Tang}
\affiliation{Nano-Bio Spectroscopy group, Dpto.~F\'isica de Materiales, Universidad del Pa\'is Vasco, Centro de F\'isica de Materiales CSIC-UPV/EHU-MPC and DIPC, Av.~Tolosa 72, E-20018 San Sebasti\'an, Spain}
\affiliation{Department of Physics and State Key Laboratory of Low-Dimensional Quantum Physics, Tsinghua University, Beijing 100084, People's Republic of China}
\affiliation{Institute for Advanced Study, Tsinghua University, Beijing 100084, People's Republic of China}
\author{A. Iacomino}
\affiliation{Nano-Bio Spectroscopy group, Dpto.~F\'isica de Materiales, Universidad del Pa\'is Vasco, Centro de F\'isica de Materiales CSIC-UPV/EHU-MPC and DIPC, Av.~Tolosa 72, E-20018 San Sebasti\'an, Spain}
\author{W. Duan}
\affiliation{Department of Physics and State Key Laboratory of Low-Dimensional Quantum Physics, Tsinghua University, Beijing 100084, People's Republic of China}
\affiliation{Institute for Advanced Study, Tsinghua University, Beijing 100084, People's Republic of China}
\author{G. Merino}
\affiliation{Departamento de Fisica Aplicada, Centro de Investigacion y de Estudios Avanzados, Unidad Merida, Km. 6 Antigua Carretera a Progreso, Apdo. Postal 73, Cordemex, 97310, Merida, Yuc., Mexico}
\author{A. Rubio}\email{angel.rubio@ehu.es}
\affiliation{Nano-Bio Spectroscopy group, Dpto.~F\'isica de Materiales, Universidad del Pa\'is Vasco, Centro de F\'isica de Materiales CSIC-UPV/EHU-MPC and DIPC, Av.~Tolosa 72, E-20018 San Sebasti\'an, Spain}

\date{\today}

\begin{abstract}

In this Supplemental Material we provide a comparison between the folded band structure of 3$\times$3 silicene before and after interaction with the Ag substrate (that complement the results shown in Figs. 1 and 2 in the main text). The isosurface charge densities of certain relevant states are plotted in Fig.1. Hybridization with Ag shifts the energies of states of silicene origin down and changes their character significantly. However, enough resemblance remains such that we can make correspondence between states in the absence and presence of Ag.

\end{abstract}

\maketitle

\begin{figure}
\includegraphics[width=16cm]{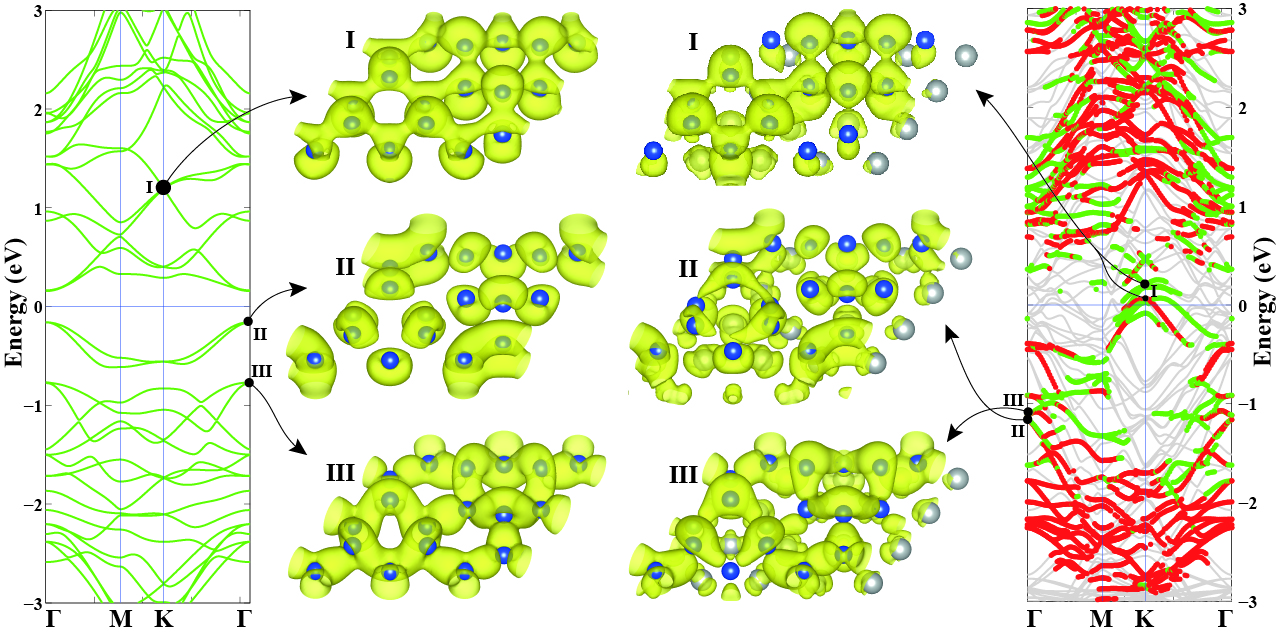}
\caption{(Color Online) The band structure (side panels) and band decomposed charge density isosurfaces (inner panels) of reconstructed silicene in the absence (left side) and presence (right side) of the Ag substrate is shown. The red color indicates the states contributed by $p_z$ orbitals of six Si atoms pointing outwards whereas the green and light grey lines correspond to the states contributed by Si and Ag atoms, respectively. This correspondence is found by calculating projections of each states to specific atomic orbitals. Si and first layer of Ag atoms are represented by blue and grey balls in the charge density plots. States of interest are labelled by Roman numerals. States corresponding to charge density isosurfaces are shown by arrows where the black circles with little, medium, and large sizes indicate the single, double, and triple degenerate states, respectively.}
\label{fig}
\end{figure}

We first look at the reconstructed silicene in the absence of the substrate which is referred as unsupported silicene. The band profile of this structure is presented in the left panel of Fig.1.  Note that, the linearly crossing bands at the K point of unreconstructed (1$\times$1) silicene is folded to the $\Gamma$-point of its (3$\times$3) supercell. Here the reconstruction breaks the symmetry and opens a 0.3~eV gap at the $\Gamma$-point. However, the linear dispersion is still present away from the $\Gamma$-point. A linear dispersion is also present in the states denoted by I whose bands cross each other at the K point. These states could be linked to six Si atoms that point outwards and form a hexagonal superstructure with three Si atoms at each corner, as seen in STM (see Fig.1 (a) in the main text). The band structure presented here is in agreement with the one reported in a very recent article.\cite{oshiyama} 

The band structure of reconstructed (3$\times$3) silicene on (4$\times$4) Ag substrate is shown in the right panel of Fig.1. To have a better insight about the role of silicene in this band structure, we computed the relative contribution of Si atoms by calculating the projection of each state on the atomic orbitals of silicene. We marked with green dots the states with a significant projection on silicene orbitals in order for the plot to preserve its clarity and information. Furthermore, the color of each green dot is changed to red if the corresponding state has significant projection on the $p_z$ orbitals of the six Si atoms positioned relatively further apart from the Ag substrate. On the base of this distinction, it is easy to see that there is a gap in the states contributed by Si atoms at the $\Gamma$-point of (3$\times$3) silicene. This is in accordance with the measured ARPES data. However, there are still states contributed by Si atoms around the Fermi level but at the K point of the (3$\times$3) silicene, thus explaining the calculated STM image at 0.0~eV (see Fig.1 (a) in the main text). Furthermore, we compare the band decomposed charge densities at K point of the (3$\times$3) reconstructed silicene with and without the Ag substrate, denoted as I in Fig.1. The similarities between these charge densities suggest that the states that contribute to the Fermi level of silicene on Ag system are coming from hybridization between Ag states and states 1.2~eV above the Fermi level of unsupported silicene. The charge transfer from Ag to silicene which is found to be only 0.5 electron per (3$\times$3) supercell cannot account for this energy shift. This issue was also discussed in Ref.\cite{oshiyama} for states at the $\Gamma$ point. Detailed charge density analysis shows that hybridization with Ag breaks the triple degenerate symmetry that this state possess in the absence of Ag.

State II corresponds to $\pi$-bands of silicene and unfolds to the K point of 1$\times$1 silicene. State III corresponds to $\sigma$-bands of silicene and remains at $\Gamma$ point after unfolding. As seen in the Fig.1, states I and II have more overlap with Ag atoms compared to state III due to their $\pi$ character. Upon hybridization, states I and II shift down by $\sim$0.9~eV while state III shifts down by $\sim$0.5~eV. This difference inverts the band order between states II and III.